\documentclass[10pt,aps,twocolumn,superscriptaddress,nofootinbib,longbibliography]{revtex4-2}
\usepackage[utf8]{inputenc}
\usepackage[T1]{fontenc}
\usepackage{braket}
\usepackage{amsfonts, amsmath, amssymb, mathtools}
\usepackage{dcolumn}
\usepackage{booktabs}
\usepackage{siunitx}
\usepackage{algorithm}
\usepackage{algpseudocode}
\usepackage{xurl} 
\usepackage[breaklinks=true,colorlinks,citecolor=blue,linkcolor=blue,urlcolor=blue]{hyperref}
\usepackage{orcidlink}
\usepackage{xcolor}
\usepackage{graphicx}
\usepackage{float}
\usepackage{comment}
\usepackage{microtype}

\begin{document}

\title{Quantum Combinatorial Reasoning for Large Language Models}

\author{Carlos Flores-Garrig\'os$^{\orcidlink{0009-0000-9735-5411}}$}
\email{carlos.flores@kipu-quantum.com}
\affiliation{Kipu Quantum GmbH, Greifswalderstrasse 212, 10405 Berlin, Germany}
\affiliation{IDAL, Electronic Engineering Department, ETSE-UV, University of Valencia, Avgda. Universitat s/n, 46100 Burjassot, Valencia, Spain}
\author{Gaurav Dev}
\affiliation{TUM School of Management, Technische Universität München, Bildungscampus 9, 74076 Heilbronn, Germany}
\author{Michael Falkenthal$^{\orcidlink{0000-0001-7802-1395}}$}
\affiliation{Kipu Quantum GmbH, Greifswalderstrasse 212, 10405 Berlin, Germany}
\author{Alejandro Gomez Cadavid$^{\orcidlink{0000-0003-3271-4684}}$}
\affiliation{Kipu Quantum GmbH, Greifswalderstrasse 212, 10405 Berlin, Germany}
\affiliation{Department of Physical Chemistry, University of the Basque Country UPV/EHU, Apartado 644, 48080 Bilbao, Spain}
\author{Anton Simen$^{\orcidlink{0000-0001-8863-4806}}$}
\affiliation{Kipu Quantum GmbH, Greifswalderstrasse 212, 10405 Berlin, Germany}
\affiliation{Department of Physical Chemistry, University of the Basque Country UPV/EHU, Apartado 644, 48080 Bilbao, Spain}
\author{Shubham Kumar$^{\orcidlink{0000-0003-2891-1669}}$}
\affiliation{Kipu Quantum GmbH, Greifswalderstrasse 212, 10405 Berlin, Germany}
\author{Enrique Solano$^{\orcidlink{0000-0002-8602-1181}}$}
\email{enr.solano@gmail.com}
\affiliation{Kipu Quantum GmbH, Greifswalderstrasse 212, 10405 Berlin, Germany}

\author{Narendra N. Hegade}
\email{narendrahegade5@gmail.com}
\affiliation{Kipu Quantum GmbH, Greifswalderstrasse 212, 10405 Berlin, Germany}

\begin{abstract}
We design and implement a quantum combinatorial reasoning framework for large language models (QCR-LLM), integrating a real quantum computer in the hybrid workflow. QCR-LLM reformulates reasoning aggregation as a higher-order unconstrained binary optimization (HUBO) problem. In this sense, reasoning fragments are represented as binary variables and their interactions encode statistical relevance, logical coherence, and semantic redundancy. We tackle the resulting high-order optimization problem both classically, via simulated annealing, and quantumly through the bias-field digitized counterdiabatic quantum optimizer (BF-DCQO) executed on IBM's superconducting digital quantum processors. Experiments on BIG-Bench Extra Hard (BBEH) benchmarks demonstrate that our QCR-LLM consistently improves reasoning accuracy across multiple LLM backbones, surpassing reasoning-native systems such as o3-high and DeepSeek R1 by up to $+9\,$pp. Despite requiring multiple reasoning samples per query, our QCR-LLM remains approximately five times more energy-efficient than o3-high, owing to the low per-token energy footprint of its GPT-4o backbone. These results constitute the first experimental evidence of quantum-assisted reasoning, showing that hybrid quantum–classical optimization can efficiently enhance reasoning coherence, interpretability, and sustainability in large-scale language models. We have opened the doors to the emergence of quantum intelligence, where harder prompts require quantum optimizers at quantum-advantage level.

\end{abstract}

\maketitle

\section{Introduction}\label{sec:introduction}
Large Language Models (LLMs) have achieved extraordinary performance across a wide range of cognitive tasks \cite{zhao2023survey}, from language understanding \cite{hoscilowicz2024llmExpansion} and translation to reasoning \cite{wei2022chain, kojima2022large} and planning as agents \cite{luo2025largelanguagemodelagent, team2023gemini, yao2022react}. Their success relies not only on scale but also on improved prompting strategies that enable them to articulate intermediate steps —a capability often referred to as reasoning \cite{srivastava2022beyond}. Among these, the Chain-of-Thought (CoT) paradigm \cite{wei2022chain} has become the de facto mechanism for improving interpretability and factual consistency in complex problem-solving. By prompting a model to think step-by-step, CoT reveals the latent reasoning pathways embedded in the network's internal representations.

However, despite its effectiveness, CoT reasoning is not without limitations. Independent CoT samples often contain redundant or contradictory reasoning fragments, unstable logical flow, or hallucinated premises. These issues are particularly prominent in open-ended or multi-hop reasoning tasks. Recent variants of CoT prompting—such as Self-Consistency \cite{wang2022self}, Tree-of-Thoughts (ToT)\cite{yao2023tree}, and Voting-based Reasoning \cite{wang2025ranked}—attempt to address these limitations by aggregating multiple reasoning paths, exploring diverse trajectories, or selecting the most consistent final answers. Although ToT and self-consistency represent major progress toward structured reasoning aggregation, they still rely on heuristic voting or traversal mechanisms that treat reasoning trajectories as discrete textual objects, without explicitly modeling their underlying semantic dependencies. 

Recent work has introduced the notion of Combinatorial Reasoning, where reasoning aggregation is formulated as a Quadratic Unconstrained Binary Optimization (QUBO) problem \cite{esencan2024combinatorial}. In this framework, each reasoning fragment is represented as a binary variable, and the objective function encodes the importance and compatibility of fragment pairs through frequency and co-occurrence statistics. The optimization process seeks the combination of fragments that minimizes the total energy, effectively selecting the most consistent reasoning chain. Although this formulation successfully captures low-order dependencies between reasoning elements, it remains restricted to quadratic (two-body) interactions. As reasoning complexity increases, coherence often depends on collective three-body and higher-order relationships that must be modeled, and representing these within a QUBO requires introducing auxiliary variables that grow rapidly in number, inflating the formulation and making such reductions impractical \cite{Jun_2023}.

To overcome these limitations, we extend this framework to the Higher-Order Unconstrained Binary Optimization (HUBO) domain. In our formulation, reasoning aggregation is modeled with an energy function that includes explicit $k$-body interaction terms to capture multi-fragment dependencies, logical coherence, and semantic redundancy at once. This generalization yields a richer, higher-dimensional landscape in which each coefficient reflects statistical signals such as frequency and co-occurrence, as well as semantic similarity across fragments. We integrate similarity-aware penalties directly into the HUBO coefficients, thereby discouraging redundant fragments while favoring complementary reasons.

However, this expressive power comes with computational cost: including $k$-body interactions causes the number of distinct terms to grow rapidly with interaction order and problem size, inflating the search space and stressing classical heuristics. This motivates the use of quantum optimization in precisely this regime. To explore these rugged, high-dimensional energy landscapes more efficiently, we employ digital quantum solvers such as the BF-DCQO)~\cite{Cadavid2025bias, romero2025bias}, which can tackle HUBO instances with $k$-local interactions directly on current digital quantum hardware without quadratic reduction. BF-DCQO uses engineered bias fields in an iterative schedule together with digitized counterdiabatic protocols~\cite{hegade2021shortcuts, hegade2022digitized} to traverse complex energy landscapes and recover near-optimal solutions that are difficult to obtain with classical optimization alone~\cite{chandarana2025runtime}.

The remainder of the paper is organized as follows. Section~\ref{sec:pipeline} introduces the multi-sample CoT pipeline and its mapping to a general $k$-body HUBO. Section~\ref{sec:solvers} presents the classical and quantum solvers used for optimization. Section~\ref{sec:pipeline_full} assembles the end-to-end workflow, from multi-sample CoT to the final prompt. Section~\ref{sec:experiments} reports results on BIG-Bench Extra Hard (BBEH)~\cite{kazemi2025big}. Finally, we outline future directions toward order-aware, quantum-accelerated reasoning and discuss implications for scalable AI reasoning architectures.

\section{From Multi-Sample chains to HUBO}
\label{sec:pipeline}

Building on the need to capture collective, higher-order relationships among reasoning fragments, we present the data-to-model pipeline that maps multi-sample CoT outputs to a HUBO instance. The procedure consolidates raw traces into a normalized fragment pool and assigns binary variables that support $k$-body interactions for downstream optimization.

For each question $q$, we begin by generating $N$ independent zero-shot completions from one or several LLMs, using fixed decoding parameters to ensure comparability across samples. Each completion produces a structured reasoning trace composed of short, self-contained fragments that we denote as \emph{reasons}. Using sentence embeddings, these fragments are extracted, cleaned, and semantically normalized to remove redundancies or stylistic noise. Then, we compute pairwise cosine similarities between all fragments and merge those whose semantic distance falls below a predefined threshold, resulting in a consolidated set of $R$ distinct reasoning fragments $\{r_i\}_{i=1}^R$. This step transforms the raw model output into a normalized reasoning pool that serves as the basis for combinatorial selection.

Each reasoning fragment $r_i$ is assigned a binary decision variable $x_i \in \{0,1\}$, indicating whether the fragment is included in the final aggregated reasoning sequence ($x_i=1$) or not ($x_i=0$). Equivalently, we can express these variables as Ising spins $z_i \in \{-1,+1\}$ via the transformation $x_i = \tfrac{1-z_i}{2}$, which facilitates later mapping to quantum hardware. The objective of our framework is to find the configuration $\mathbf{x}$ (or $\mathbf{z}$) that minimizes an energy function that represents the global coherence, diversity and statistical relevance of the selected fragments.

Reasoning aggregation is formalized as a HUBO problem, whose energy function takes the general form
\begin{equation}
\label{eq:kbody}
H(\mathbf{x}) = \sum_{\emptyset\neq S\subseteq [R],\ |S|\le K} w_S \prod_{i\in S} x_i,
\end{equation}
where each coefficient $w_S \in \mathbb{R}$ encodes the statistical and semantic relationships among the subset of fragments indexed by $S$. The order of interaction $|S|$ determines the locality of the term: 1-body terms correspond to individual fragment properties, 2-body terms encode pairwise relations, and 3-body or higher-order terms capture collective dependencies among multiple fragments. Although our current implementation employs $K \in \{2,3\}$, the formulation generally generalizes to arbitrary $K$.

\subsection{Coefficient Design}

The construction of coefficients $w_S$ integrates both statistical and semantic information extracted from the multi-sample reasoning set. The linear (1-body) terms quantify the intrinsic importance and stability of each fragment. For each $r_i$, we estimate its empirical popularity $p_i = \tfrac{n_i}{N}$, defined as the fraction of completions in which the fragment (or a close semantic variant) appears. The variability of its occurrence across completions, $\mathrm{risk}_i = p_i (1 - p_i)$, measures its stability: fragments that are either too common or too rare provide little discriminative information. Combining both factors, we define the linear coefficients as
\begin{equation}
\label{eq:linear}
w_i = -\mu\, p_i + \alpha\, \mathrm{risk}_i,
\end{equation}
where $\mu$ and $\alpha$ are positive hyperparameters controlling the trade-off between representativeness and variability. High-popularity fragments contribute negatively to the energy (and are therefore favored), whereas unstable ones receive a positive penalty.

Pairwise relations between fragments are encoded through the quadratic (2-body) coefficients. For each pair $(i,j)$, we compute their connected correlation
\begin{equation}
\label{eq:cij}
c_{ij} = \frac{n_{ij}}{N} - p_i p_j,
\end{equation}
which measures whether fragments $r_i$ and $r_j$ tend to co-occur more or less frequently than expected under independence. After standardizing these correlations into $\tilde{c}_{ij}$ (e.g., z-scores with regularization $\varepsilon$), we assign
\begin{equation}
w_{ij} = -\beta\, \big(\tilde{c}_{ij} - \lambda^{(2)}_{\mathrm{sim}}\, \mathrm{sim}(i,j)\big),
\end{equation}
where $\beta$ is a scaling factor and $\lambda^{(2)}_{\mathrm{sim}}$ controls the penalty for semantic redundancy based on cosine similarity $\mathrm{sim}(i,j)$. Thus, pairs that co-occur frequently but express distinct ideas are energetically rewarded, while highly similar or redundant pairs are penalized.

To capture higher-order dependencies, we extend this formulation to triplets and general $k$-body interactions. For a triplet $(i,j,k)$, we define a connected three-body correlation
\begin{equation}
c_{ijk} = \frac{n_{ijk}}{N} - p_i p_j p_k,
\end{equation}
and its corresponding coefficient
\begin{equation}
w_{ijk} = -\gamma\, \big(\tilde{c}_{ijk} - \lambda^{(3)}_{\mathrm{sim}}\, \overline{\mathrm{sim}}(i,j,k)\big),
\end{equation}
where $\tilde{c}_{ijk}$ is the normalized 3-body correlation, $\gamma$ controls its contribution, and $\overline{\mathrm{sim}}(i,j,k)$ is the mean pairwise similarity among the three fragments. This term rewards sets of fragments that tend to co-occur coherently while maintaining semantic diversity. The same principle generalizes to arbitrary $k$ by replacing $\overline{\mathrm{sim}}(i,j,k)$ with a symmetry-consistent similarity aggregation $\overline{\mathrm{sim}}_S$ for each subset $S$.

After computing all coefficients, we normalize them type-wise to ensure numerical stability and compatibility with both classical and quantum solvers. Specifically, 1-body, 2-body, and 3-body coefficients are rescaled into distinct ranges $[-a,a]$, $[-b,b]$, and $[-c,c]$, preserving contrast within each order of interaction. This normalization avoids dominance of any interaction order and ensures that the optimization landscape remains well-conditioned.

The resulting energy function $H(\mathbf{x})$ defines a structured Hamiltonian over the reasoning fragments, where low-energy configurations correspond to coherent, diverse, and semantically consistent combinations of reasons. For $K=3$, the explicit form of the Hamiltonian is
\begin{equation}
H(\mathbf{x}) = \sum_i w_i x_i + \sum_{i<j} w_{ij} x_i x_j + \sum_{i<j<k} w_{ijk} x_i x_j x_k,
\end{equation}
with coefficients $w_i$, $w_{ij}$, and $w_{ijk}$ derived as described above. This $3$-body Hamiltonian captures both statistical co-occurrence and semantic structure across reasoning fragments, forming the foundation upon which the optimization and quantum-solving procedures operate in subsequent sections.

% ================================================================
\section{Solvers and Reasoning Selection}
\label{sec:solvers}

The optimization landscape defined by the HUBO formulation can be explored through both classical and quantum algorithms, depending on the problem size and the order of interactions involved. For lower-order or sparse Hamiltonians ($K \le 3$), classical methods remain a practical option, while for denser and higher-order configurations, quantum solvers have the potential to provide computational advantage \cite{chandarana2025runtime}. 

In the classical regime, we employ SA~\cite{kirkpatrick1983optimization} as a baseline optimizer. Although SA can, in principle, handle HUBO formulations directly, we restrict our implementation to QUBO instances since we use the D-Wave Neal solver, which operates on quadratic forms. Therefore, when $K \le 3$, the HUBO is approximately reduced to a QUBO by redistributing cubic contributions among pairwise couplings, yielding an effective formulation that preserves the main structural correlations of the reasoning landscape. SA minimizes the energy by probabilistically accepting or rejecting spin flips according to a temperature schedule, progressively converging toward configurations of minimal energy. By running multiple annealing trajectories, we obtain not only a candidate ground state but also a distribution of near-optimal solutions that collectively describe the low-energy manifold of the reasoning problem.

As the order of interactions increases, the HUBO landscape becomes highly non-convex and frustrated, rendering classical approaches exponentially inefficient. To address this, we implement the BF-DCQO) \cite{romero2025bias, Cadavid2025bias} on IBM digital quantum backends. The HUBO model is first mapped to an equivalent spin Hamiltonian via the transformation $x_i = (1 - z_i)/2$, where $z_i \in \{-1,+1\}$ represents the spin state associated with each reasoning fragment $r_i$. The resulting Hamiltonian is encoded onto a system of qubits whose couplings reproduce the $k$-body interactions defined in the model. Unlike classical approximations that require quadratic reduction, BF-DCQO handles higher-order couplings natively by introducing engineered bias fields and counterdiabatic driving schedules. The algorithm evolves the system toward its ground-state manifold, and repeated quantum executions (shots) generate a probability distribution over bitstrings that sample the low-energy configurations of the Hamiltonian. This output histogram provides both the most probable bitstring (ground-state candidate) and the marginal inclusion frequencies of each qubit, interpreted as the selection probabilities of the corresponding reasoning fragments.

\subsection{Reason Selection and Stability Ranking}

\begin{figure*}[t]
    \centering
    % ---- Panel (a): Energy distribution  ----
    \includegraphics[width=\linewidth]{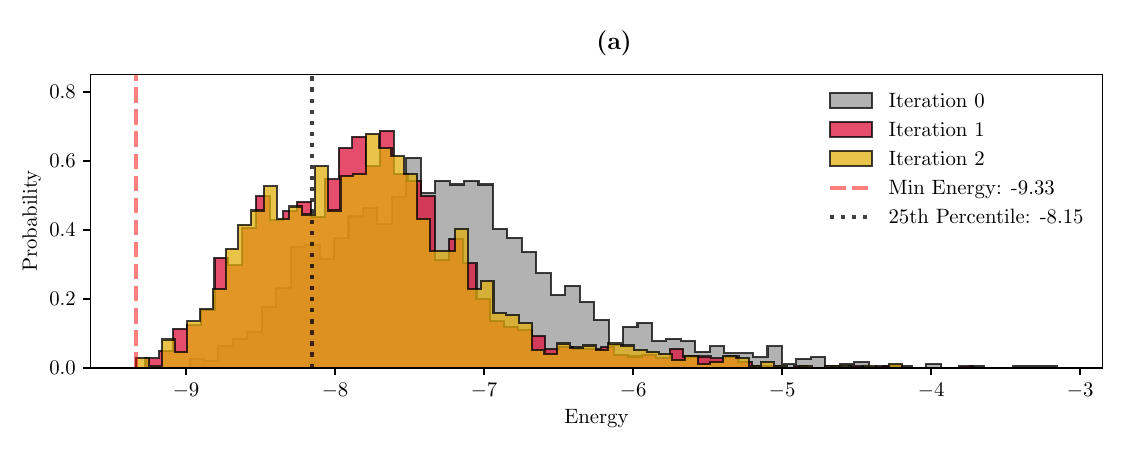}
    % ---- Panel (b): Feature selection  ----
    \includegraphics[width=\linewidth]{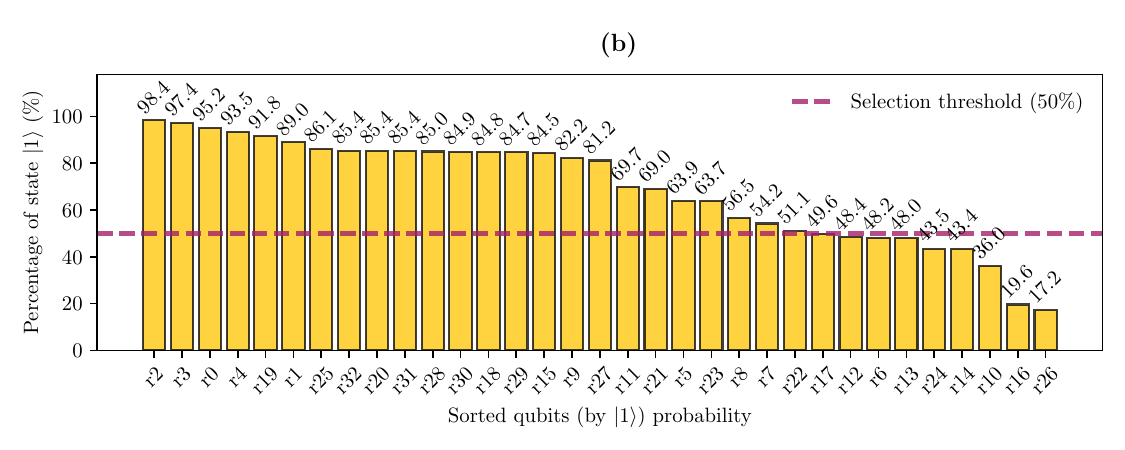}
    \vspace{0.4em}
    \caption{\textbf{Example of the BF-DCQO optimization process for a single question from the \textit{Causal Understanding} dataset.}
    \textbf{(a)} \textbf{Energy distributions obtained over three BF-DCQO iterations (\textit{ibm-aachen})}. 
    The histogram shows convergence of the energy landscape toward lower minima; 
    the dashed red and dotted black lines indicate respectively the minimum energy 
    and the 25th percentile threshold used to define the low-energy subset. 
    \textbf{(b)} \textbf{Expected inclusion frequencies of each reason computed 
    within this 25\% lowest-energy set}. The dashed magenta line represents the 
    adjustable stability threshold: increasing it yields fewer but more consistent 
    reasons, while relaxing it broadens the reasoning basis. 
    Together, both panels illustrate how the BF-DCQO solver provides an interpretable 
    energy landscape from which stable reasoning fragments are ranked and selected.}
    \label{fig:bf_dcqo}
\end{figure*}

Once the optimization process, classical or quantum, has been executed, we analyze the resulting ensemble of solutions to extract interpretable reasoning information. Rather than relying solely on the single best configuration, we focus on the statistical structure of low-energy solutions, which provides a more robust signal of reasoning stability. To this end, we collect the lowest 25\% of energy configurations from the full sample of annealing runs or quantum measurement outcomes. For each fragment $r_i$, we compute its empirical inclusion frequency within this subset, corresponding to the expected value $\langle x_i \rangle$ (or equivalently, $\tfrac{1-\langle z_i \rangle}{2}$ in spin representation).

Fragments that appear consistently across these near-optimal configurations are considered highly stable and form the backbone of the aggregated reasoning chain, while those with fluctuating or marginal inclusion are treated as context-dependent or peripheral. This frequency-based ranking naturally differentiates essential reasoning steps from optional or redundant ones, providing both interpretability and quantitative insight into the reasoning process. The final reasoning subset can thus be obtained by thresholding the selection frequencies or by taking the ground-state configuration when a single deterministic reasoning chain is desired.

This two-stage analysis—optimization followed by statistical ranking—transforms the raw output of combinatorial optimization into an interpretable reasoning hierarchy. It bridges the gap between symbolic reasoning and probabilistic aggregation, while leveraging the computational advantages of quantum solvers to explore reasoning landscapes that are intractable by classical means.

As shown in Fig.~\ref{fig:bf_dcqo}, the BF-DCQO solver yields an interpretable energy landscape whose low-energy configurations define the stable reasoning subset. In this example, we illustrate how QCR-LLM selects reasoning fragments for a single question that initially produced 33 distinct reasons across the multi-sample completions. After optimization, only 24 of these reasons remain in the final subset, corresponding to those with the highest expected inclusion frequencies within the 25\% lowest-energy configurations. The selection threshold is here set at 50\% appearance probability, meaning that a reason must appear in at least half of the low-energy solutions to be retained. This threshold, however, can be adjusted to control the trade-off between informational richness and prompt compactness: a lower threshold includes more reasons, increasing token count and contextual detail, while a higher threshold yields a more concise yet semantically diverse reasoning chain with reduced token usage. This flexibility allows practitioners to tune QCR-LLM depending on whether interpretability, efficiency, or completeness is prioritized in the final reasoning output.

\section{End-to-End Quantum Combinatorial Reasoning}
\label{sec:pipeline_full}

\begin{figure*}[t]
    \centering
    \includegraphics[height=8.5cm,width=16cm]{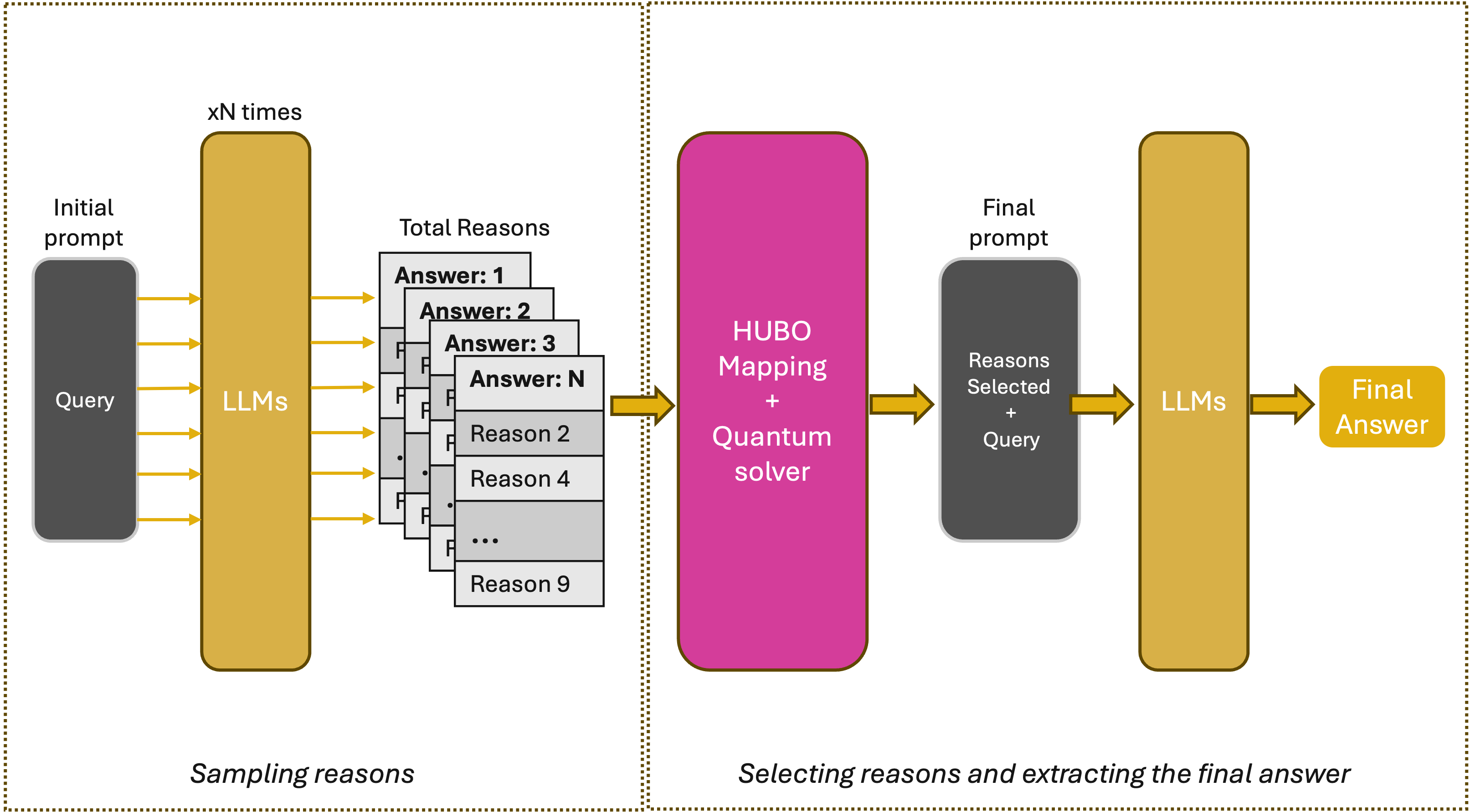}
    \caption{\textbf{Overview of the Quantum Combinatorial Reasoning (QCR-LLM) pipeline.}
    Multiple zero-shot Chain-of-Thought completions are sampled from the base LLM, producing independent reasoning trajectories.
    Distinct reasoning fragments (reasons) are extracted and aggregated into a high-order binary optimization model (HUBO).
    The resulting Hamiltonian is solved via the Bias-field Digitized Counterdiabatic Quantum Optimization (BF-DCQO) on IBM digital quantum hardware.
    The selected reasoning subset is then reintroduced into the model as contextual evidence to produce the final answer.}
    \label{fig:qcr_pipeline}
\end{figure*}

The complete Quantum Combinatorial Reasoning (QCR-LLM) process integrates the methodological components described in the previous sections into a unified reasoning pipeline, as illustrated in Fig.~\ref{fig:qcr_pipeline}. Given an input query, we first perform multi-sample zero-shot Chain-of-Thought prompting using the base LLM, generating $N$ independent reasoning traces under identical decoding conditions. Each trace produces multiple atomic fragments, or reasons, which are parsed, normalized, and semantically deduplicated using the embedding-based similarity functions described in Sec.~\ref{sec:pipeline}. The resulting set of distinct reasons $\{r_i\}_{i=1}^R$ constitutes the combinatorial basis for the optimization stage. 

These fragments are then mapped into a structured HUBO model following Eq.~(\ref{eq:kbody}), where the coefficients $w_S$ are computed as detailed in Sec.~\ref{sec:pipeline}: individual importance and risk through Eq.~(\ref{eq:linear}), pairwise correlations via Eq.~(\ref{eq:cij}), and higher-order dependencies incorporating semantic redundancy penalties. The resulting Hamiltonian encodes both statistical and semantic relations among reasoning fragments, forming a high-dimensional energy landscape whose minima correspond to coherent and diverse reasoning subsets.

The HUBO is then solved using the optimization procedures introduced in Sec.~\ref{sec:solvers}. For small or sparse instances, we can apply classical SA, while dense or high-order configurations are addressed using the Bias-field Digitized Counterdiabatic Quantum Optimization (BF-DCQO) implemented on IBM digital quantum hardware. The BF-DCQO solver exploits engineered bias fields and counterdiabatic schedules to efficiently traverse the low-energy manifold of the Hamiltonian, directly handling $k$-body interactions without quadratic reduction. The output here consists of both the ground-state bitstring and a distribution of near-optimal configurations from which the expected inclusion $\langle x_i \rangle$ of each reason is estimated.

Finally, using the stability ranking procedure outlined in Sec.~\ref{sec:solvers}, we identify the subset of reasoning fragments that consistently appear across the lowest-energy configurations. These stable reasons are reintroduced into the LLM as contextual evidence, together with the original query, forming a final structured prompt that guides the model to generate the final answer. This last step closes the reasoning loop, combining quantum-optimized structural selection with linguistic inference. In doing so, QCR-LLM transforms the inherently stochastic reasoning behavior of LLMs into a controlled, interpretable, and physically grounded reasoning process that bridges classical statistics, semantic representation, and quantum optimization.

\begin{algorithm}[H]
\caption{Quantum Combinatorial Reasoning (QCR-LLM) Pipeline}
\label{alg:qcr}
\begin{algorithmic}[1]
\State \textbf{Input:} Query $q$, number of samples $N$, LLM model $M$
\State Generate $N$ zero-shot CoT completions: $\mathcal{A} = \{a_1, a_2, ..., a_N\}$ using $M$
\State Extract atomic reasoning fragments and deduplicate semantically to obtain $\{r_i\}_{i=1}^R$
\State Compute coefficients $w_S$ for $|S| \le K$ following Sec.~\ref{sec:pipeline}
\State Build HUBO Hamiltonian $E(\mathbf{x}) = \sum_{|S|\le K} w_S \prod_{i\in S} x_i$
\State Solve using SA/BF-DCQO
\State Collect low-energy configurations and compute inclusion frequencies $\langle x_i \rangle$
\State Select stable fragments $\{r_i : \langle x_i \rangle \ge \tau\}$ or ground-state subset
\State Form final prompt $p_f = (q, \{r_i\}_{\text{stable}})$
\State Query LLM with $p_f$ to produce the \textit{Final Answer}
\end{algorithmic}
\end{algorithm}

The pseudocode in Algorithm~\ref{alg:qcr} summarizes the complete QCR-LLM pipeline, aligning with the schematic in Fig.~\ref{fig:qcr_pipeline}. It highlights the modular structure of the framework: reasoning generation, semantic aggregation, HUBO construction, solver execution (classical or quantum), and the final synthesis step that reintroduces optimized reasoning fragments into the language model to yield a refined, interpretable answer.

\section{Experiments on BIG-Bench Extra Hard Tasks}\label{sec:experiments}

To evaluate the proposed QCR-LLM framework, we benchmarked multiple configurations on three representative subsets of BBEH~\cite{kazemi2025big}: 
\textit{Causal Understanding}, \textit{DisambiguationQA}, and \textit{NYCC (New York Conceptual Combinations)}. 
These datasets were chosen for their diversity in reasoning type and output structure. 
In \textit{Causal Understanding}, each question requires determining whether a causal relationship holds, with possible answers \texttt{Yes}, \texttt{No}, or \texttt{Ambiguous}. 
\textit{DisambiguationQA} involves multi-choice semantic reasoning with options from \texttt{A} to \texttt{H} (up to eight candidate answers). 
Finally, \textit{NYCC} extends this structure further, offering up to ten multiple-choice options (\texttt{A}–\texttt{J}) that require complex conceptual blending and associative reasoning. 
Accuracy is computed as the fraction of questions for which the model selects the correct option.

\subsection{Simulated Annealing}

The results in Table~\ref{tab:bbh_qrllm_models} are obtained by optimizing the HUBO Hamiltonian using classical Simulated Annealing (SA), which serves as our primary solver for the experimental phase. 
Since the SA solver we used operates on quadratic energy forms, the original 3-body Hamiltonian

\begin{equation}
H(\mathbf{z}) = \sum_i c_i z_i + \sum_{i<j} c_{ij} z_i z_j + \sum_{i<j<k} c_{ijk} z_i z_j z_k ,
\end{equation}

must first be reduced to an equivalent quadratic formulation QUBO through the introduction of auxiliary variables. 
This transformation distributes the contribution of the cubic term $c_{ijk} z_i z_j z_k$ into several pairwise couplings while preserving the overall energy landscape:
\begin{equation}
c_{ijk} z_i z_j z_k \;\rightarrow\; 
\frac{1}{2}c_{ijk}(z_i z_j + z_i z_k + z_j z_k) - \frac{1}{2}c_{ijk}, \label{eq:hubo_to_qubo}
\end{equation}
so that the effective Hamiltonian becomes
\begin{equation}
H_{\text{QUBO}}(\mathbf{z}) =
\sum_i \tilde{c}_i z_i +
\sum_{i<j} \tilde{c}_{ij} z_i z_j,
\end{equation}
where $\tilde{c}_i$ and $\tilde{c}_{ij}$ collect both the original 2-body coefficients and the redistributed 3-body contributions. 
This reduction allows the HUBO to be handled by standard classical annealers while retaining its structural semantics.

The base models in the upper half of Table~\ref{tab:bbh_qrllm_models} correspond to leading LLMs featured in the BBEH report~\cite{kazemi2025big}. 
Each QCR-LLM variant is produced by aggregating 20 independent zero-shot Chain-of-Thought completions from the same model, followed by HUBO optimization and re-injection of the selected reasoning fragments, described in detail in Sec.~\ref{sec:pipeline_full}.  
Across all, the QCR-LLM approach consistently improves accuracy over the corresponding base model. 
For instance, QCR-LLM (GPT-4o) achieves \textbf{+5.5}~pp in Causal (59.5 vs.~54.0), \textbf{+8.3}~pp in Disambiguation (60.0 vs.~51.7), and \textbf{+1.5}~pp in NYCC (24.5 vs.~23.0).  
Similarly, QCR-LLM (DeepSeek V1) surpasses its base model by \textbf{+8.0}, \textbf{+9.0}, and \textbf{+4.5}~pp, respectively, while QR-LLM (LLaMA 3.1) shows gains of \textbf{+12.5}, \textbf{+28.3}, and \textbf{+3.5}~pp.  
These improvements confirm that the HUBO-based aggregation effectively distills the most coherent reasoning fragments, outperforming single-sample generation in every case.

When compared against reasoning-native baselines, such as DeepSeek R1 and o3-high, QCR-LLM also achieves higher accuracy across all datasets. 
QCR-LLM (GPT-4o) exceeds o3-high by \textbf{+5.5}~pp on Causal, \textbf{+1.7}~pp on Disambiguation, and \textbf{+8.5}~pp on NYCC. 
Likewise, QCR-LLM (DeepSeek V1) outperforms DeepSeek R1 by \textbf{+0.5}~pp, \textbf{+2.5}~pp, and \textbf{-2.5}~pp, respectively, indicating competitive or superior performance even against architectures natively designed for reasoning. 
Overall, in almost all cases, QCR-LLM configurations demonstrate positive transfer, confirming that quantum combinatorial selection can enhance reasoning quality without retraining or parameter updates.

Additionally, the QCR-LLM configuration explores a heterogeneous sampling strategy where the 20 reasoning completions are not drawn from a single model but from a mixture of three: 
7 samples from GPT-4o, 7 from DeepSeek V1, and 6 from LLaMA 3.1. 
After HUBO optimization, the selected fragments are used to construct the final prompt, which is then evaluated using GPT-4o. 
This cross-model aggregation leverages the complementary reasoning patterns of different architectures, achieving balanced improvements across all tasks (54.0, 55.8, and 20.5) and demonstrating that QCR-LLM can fuse heterogeneous reasoning signals into a unified, high-quality reasoning sequence.

It is worth noting that all models were accessed through the Azure OpenAI Service, which provides direct API access to GPT-4o and o3-high. 
Additional models, including LLaMA and DeepSeek families, were interfaced via \textit{LangChain} connectors using their respective hosted endpoints. 
All evaluations used identical decoding settings (temperature, top-p, and token limits) to ensure comparability across backbones. 
Classical optimization and quantum post-processing were performed locally using the QCR-LLM framework described in Sec.~\ref{sec:pipeline_full}.

\begin{table}[h!]
\centering
\caption{\textbf{Performance of base and quantum-optimized reasoning models on BIG-Bench Extra Hard tasks.} 
Each entry reports task accuracy (\%). 
Standard LLMs (top) are compared against their corresponding QCR-LLM variants (bottom), 
where multi-sample reasoning aggregation is optimized through the HUBO formulation and 
quantum solving via BF-DCQO. 
The last row (QCR-LLM Combined) aggregates reasoning fragments jointly sampled 
from GPT-4o~\cite{openai2024gpt4ocard}, DeepSeek V1, and LLaMA 3.1~\cite{grattafiori2024llama3herdmodels}, demonstrating multi-model synergy. 
The best performance per task is highlighted in bold.}
\label{tab:bbh_qrllm_models}
\scriptsize
\setlength{\tabcolsep}{3.5pt}
\renewcommand{\arraystretch}{1.15}
\begin{tabular}{lccc}
\toprule
\textbf{Model / Variant} & 
\textbf{Causal} & 
\textbf{Disambiguation} & 
\textbf{NYCC} \\
\midrule
LLaMA 3.1 & 38.0 & 21.0 & 10.0 \\
Gemma 2.0 & 37.0 & 36.7 & 13.0 \\
G. Lite 2.0 & 45.5 & 45.0 & 13.5 \\
DeepSeek V1 & 47.0 & 43.5 & 13.0 \\
G. Flash 2.0 & 52.5 & 50.0 & 13.5 \\
GPT-4o & 54.0 & 51.7 & 23.0 \\
Qwen 32b & 54.5 & 52.5 & 10.5 \\
DeepSeek R1 & 54.5 & 50.0 & 20.0 \\
o3-high & 54.0 & 58.3 & 16.0 \\
\midrule
\textbf{QCR-LLM (GPT-4o)} & \textbf{58.5} & \textbf{60.0} & \textbf{24.5} \\
\textbf{QCR-LLM (DeepSeek V1)} & 55.0 & 52.5 & 17.5 \\
\textbf{QCR-LLM (LLaMA 3.1)} & 50.5 & 49.3 & 13.5 \\
\textbf{QCR-LLM (Combined)} & 54.0 & 55.8 & 20.5 \\
\bottomrule
\end{tabular}
\end{table}

% ================================================================
\subsection{Results with Quantum Solvers (BF-DCQO)}
\label{sec:bf_dcqo_results}

Due to runtime and access constraints on quantum hardware, we restricted our quantum experiments to the best-performing configuration, QCR-LLM (GPT-4o). 
The corresponding HUBO instances were executed on the \textit{ibm-aachen} quantum backend using the BF-DCQO. 
Unlike classical SA, which requires reducing cubic interactions to quadratic form, BF-DCQO natively handles higher-order Hamiltonians, allowing direct optimization of $k$-body couplings without structural simplification.

As shown in Table~\ref{tab:bf_dcqo_results}, the BF-DCQO solver slightly improves over the classical SA baseline: \textbf{+1.0}~pp on \textit{Causal Understanding}, stable performance on \textit{DisambiguationQA}, and \textbf{+0.5}~pp on \textit{NYCC}. 
Although these gains are modest, they were obtained on real quantum hardware and without any HUBO$\rightarrow$QUBO reduction, demonstrating that the solver preserves the original high-order structure of the reasoning Hamiltonian.

The limited but positive improvement can be explained by two main factors. 
First, the HUBO instances used in this evaluation are still moderate in size—containing tens rather than hundreds of reasoning fragments—where classical annealing already approximates near-optimal solutions efficiently. 
Second, the quantum solver’s advantage lies in its ability to represent higher-order dependencies exactly: while SA requires the approximation shown in Eq.~\ref{eq:hubo_to_qubo},
BF-DCQO directly encodes $c_{ijk} z_i z_j z_k$ in the hardware through bias-field couplings and counterdiabatic driving. 
Avoiding this reduction eliminates approximation bias and maintains the full generality of the reasoning interactions.

Additionally, an analysis of the average number of unique reasoning fragments (mapped to qubits) per question shows a clear relationship between reasoning complexity and qubit count. 
For \textit{Causal Understanding}, questions contain on average $35.4$ unique reasons (median $35$, range $5$–$63$), of which about $21$ are selected. 
\textit{DisambiguationQA} requires approximately $47.8$ unique reasons (median $48$, range $24$–$67$) with $27$ selected on average, while \textit{NYCC} is the most complex, averaging $89.4$ unique reasons (median $90$, range $5$–$120$) with roughly $22$ selected. 
This correlation between the number of qubits and task difficulty aligns with the observations in the BBEH benchmark: tasks demanding broader conceptual integration tend to generate richer and more entangled reasoning structures.
\begin{table}[h!]
\centering
\caption{\textbf{Performance comparison between classical (SA) and quantum (BF-DCQO) solvers for QCR-LLM (GPT-4o).} 
All values report task accuracy (\%).}
\label{tab:bf_dcqo_results}
\scriptsize
\resizebox{0.996\linewidth}{!}{%
\setlength{\tabcolsep}{4pt}
\renewcommand{\arraystretch}{1.15}
\begin{tabular}{lccc}
\toprule
\textbf{Solver} & \textbf{Causal} & \textbf{Disambiguation} & \textbf{NYCC} \\
\midrule
QCR-LLM SA (GPT-4o) & 58.5 & \textbf{60.0} & 24.5 \\
QCR-LLM BF-DCQO (GPT-4o) & \textbf{59.5} & \textbf{60.0} & \textbf{25.0} \\
\bottomrule
\end{tabular}
}
\end{table}

These findings suggest that the potential benefits of quantum solvers will become more pronounced as reasoning tasks grow in combinatorial complexity. 
Classical approaches remain competitive for small to mid-scale HUBOs but scale poorly with the exponential growth of $k$-body terms.  
Although current hardware limits the number of qubits and the circuit depth available, this experiment on \textit{ibm-aachen} provides a first empirical indication that direct high-order quantum optimization can enhance reasoning stability and accuracy in realistic language-model settings.

\subsection{Comparative Energy Consumption}
\label{self:Energy_Consumption}

Understanding the energy footprint of inference is essential for assessing the sustainability of large language models (LLMs). Following the methodology described by Oviedo et~al.~\cite{oviedo2025energyuseaiinference}, we express energy consumption as watt-hours per token, normalized for inference on NVIDIA H100 GPUs (700\,W TDP, $\sim$1.5\,kW per node under full load) with typical utilization levels of 10\% compute efficiency and 70\% power utilization.

This configuration reflects large-scale, production-grade deployments used by OpenAI, Meta and DeepSeek. For a standard inference window of 500 tokens in total (input + output)—the maximum size reached in our experiments—the median energy consumption for \textit{GPT-4o}~\cite{openai2024gpt4ocard} is approximately 0.3\,Wh per query ($3\times10^{-4}$\,Wh/token), in agreement with Microsoft’s large-scale inference benchmarking~\cite{oviedo2025energyuseaiinference}. By contrast, the reasoning-intensive \textit{o3-high}~\cite{openai2024openaio1card} model, which generates significantly longer chains-of-thought, consumes over 33\,Wh per long prompt ($3.3\times10^{-2}$\,Wh/token), representing an energy cost roughly two orders of magnitude higher. This difference is primarily due to extended decoding lengths and higher active parameter utilization during reasoning.

Other state-of-the-art systems, including Meta’s LLaMA~3.1~\cite{grattafiori2024llama3herdmodels} and DeepSeek’s V1 and R1 models, exhibit intermediate energy costs. Measurements reported in~\cite{oviedo2025energyuseaiinference} indicate 0.43\,Wh per query for LLaMA~3.1~405B ($8.6\times10^{-4}$\,Wh/token) and 0.5–2.25\,Wh for DeepSeekV1/R1~\cite{deepseekai2025deepseekr1incentivizingreasoningcapability} models depending on quantization and throughput optimization. These models therefore fall between GPT-4o’s efficiency regime and the high reasoning-cost regime of o3-high. Table~\ref{tab:energy_models_extended} summarizes the approximate per-token energy used for all models considered in this study.

\begin{table*}[t!]
\centering
\caption{\textbf{Estimated energy consumption per token across major LLMs.} 
Values correspond to typical inference workloads on NVIDIA H100 hardware. 
Reported values are approximate Wh/token averages derived from large-scale benchmarking~\cite{oviedo2025energyuseaiinference}.}
\label{tab:energy_models_extended}
\scriptsize
\setlength{\tabcolsep}{8pt}
\renewcommand{\arraystretch}{1.15}
\begin{tabular}{lcc}
\toprule
\textbf{Model} & \textbf{Hardware (GPU configuration)} & \textbf{Energy per token (Wh/token)} \\
\midrule
GPT-4o & NVIDIA H100 (700\,W, 10\% util.) & $3.0\times10^{-4}$ \\
o3-high & NVIDIA H100 (reasoning mode) & $3.3\times10^{-2}$ \\
LLaMA~3.1 (405B) & NVIDIA H100 (FP8 quant.) & $8.6\times10^{-4}$ \\
DeepSeek~V1 & NVIDIA H100 (FP8 quant.) & $1.0\times10^{-3}$–$2.0\times10^{-3}$ \\
DeepSeek~R1 & NVIDIA H100 (reasoning, 10×H100 cluster) & $4.5\times10^{-3}$ \\
\bottomrule
\end{tabular}
\end{table*}

\par\vspace{6pt}
This confirms that, under realistic inference workloads capped at 500 tokens, the total energy footprint per query in our experiments remains within the sub–watt-hour range. However, reasoning-extended models such as o3-high or DeepSeek-R1 can increase energy costs by more than two orders of magnitude when output lengths reach several thousand tokens, highlighting the importance of efficient token management and serving optimization in scaling sustainable AI systems.

We report LLM-side energy only and do not include the energy used by the quantum backend. A fair system-level comparison is nontrivial because cryogenic and control overheads dominate and are shared across jobs. For context, superconducting platforms can draw on the order of $10$–$25$\,kW for cryogenics plus several kilowatts for control electronics, so the per-job energy depends on utilization and amortization. A full accounting is left to future work.

% ======================================
\section{Conclusion}
\label{sec:conclusion}

We have introduced a quantum combinatorial reasoning large language model  (QCR-LLM), the first framework that integrates LLMs reasoning with both classical and quantum optimization. 
By reformulating reasoning aggregation as a HUBO problem, QCR-LLM encodes statistical relevance, logical coherence, and semantic redundancy through multi-body interactions. 
This formulation is solver-agnostic—compatible with classical simulated annealing and quantum algorithms such as the BF-DCQO—and has been successfully executed on real quantum hardware (\textit{ibm-aachen}), representing the first experimental demonstration of quantum-assisted reasoning for LLMs.

Empirical results on BIG-Bench Extra Hard benchmarks show that QCR-LLM consistently improves reasoning accuracy across all tested backbones, even surpassing reasoning-native models such as o3-high and DeepSeek~R1. 
These results confirm that reasoning can be treated as a structured optimization process rather than a purely linguistic one, enabling coherent, interpretable reasoning without retraining or architectural modification. 
Moreover, QCR-LLM remains model-agnostic, operating purely on reasoning fragments, and can therefore be applied to any current or future LLM, as well as to hybrid or multi-model configurations where reasoning samples are combined across architectures.

From an efficiency perspective, QCR-LLM remains energetically favorable even when accounting for its sampling overhead. 
As reported in Table~\ref{tab:energy_models_extended}, o3-high consumes $\sim\!100\times$ more energy per token than GPT-4o ($3.3\times10^{-2}$ vs.\ $3\times10^{-4}$ Wh/token). 
Although QR-LLM (GPT-4o) issues $N{=}20$ independent completions—i.e., $\sim\!20\times$ the tokens of a single GPT-4o call—its net energy per question remains $\approx\!20/100 = 0.2$ of o3-high, that is, about 5$\times$ lower, under our capped token budgets. 
Thus, QCR-LLM attains competitive, often superior, reasoning accuracy at a fraction of the energy of reasoning-native models.

Looking forward, extending QCR-LLM to interactions beyond three-local terms will allow the exploration of regimes where classical solvers begin to collapse while quantum solvers remain efficient, opening a path toward measurable quantum advantage for harder prompts. Future work will also focus on sequential and hierarchical reasoning pipelines, as well as hybrid approaches that combine reasoning samples from multiple LLMs within a unified quantum-optimized framework.

In summary, QCR-LLMs establish a first experimentally-tested paradigm for quantum-enhanced reasoning, showing that quantum algorithms can already contribute to more accurate, interpretable, and energy-efficient reasoning systems—laying the groundwork for the next generation of hybrid quantum–classical intelligence at the quantum-advantage level.

\section*{Acknowledgments}

The authors gratefully acknowledge Alan Ho, Mert Esencan, Davide Venturelli, Christoph Krieger, Lucas Harzenetter, Sebastian Romero, Shubham Kumar, and Jan Trautmann for inspirational discussions and valuable feedback that helped improve the clarity and precision of this manuscript. Their careful reading and constructive suggestions greatly contributed to refining the presentation of this work.

\bibliography{references}

\clearpage
\onecolumngrid
\appendix

\end{document}